\documentclass[aps,prd,superscriptaddress,floatfix,twocolumn]{revtex4-1}
\usepackage{graphicx}
\usepackage{amsmath}
\usepackage{amssymb}
\usepackage{hyperref}

\newcommand{\hx}{\hat x}
\newcommand{\hm}{\hat m}
\newcommand{\hr}{\hat\rho}
\newcommand{\hs}{\hat\Sigma}
\newcommand{\hz}{\hat Z}
\renewcommand\phi\varphi
\newcommand\q{q}
\renewcommand\d{d}
\newcommand\zm{\text{zm}}
\newcommand\nz{\text{nz}}
\DeclareMathOperator\sign{sign}
\DeclareMathOperator\tr{tr}

\begin{document}

\title{\boldmath Dirac spectrum of one-flavor QCD at $\theta = 0$ and
  continuity of the chiral condensate }

\author{J.J.M.~Verbaarschot}  
\affiliation{ Department of Physics and Astronomy, SUNY, Stony Brook,
  New York 11794, USA}
\author{T.~Wettig} 
\affiliation{Department of Physics, University of Regensburg, 93040
  Regensburg, Germany}

\date{\today}

\begin{abstract}
  We derive exact analytical expressions for the spectral density of   the Dirac operator at fixed $\theta$-angle in the microscopic domain   of one-flavor QCD.  These results are obtained by performing the sum   over topological sectors using novel identities involving sums of   products of Bessel functions.  Because the fermion determinant is   not positive definite for negative quark mass, the usual   Banks-Casher relation is not valid and has to be replaced by a   different mechanism first observed for QCD at nonzero chemical   potential.  Using the exact results for the spectral density we   explain how this mechanism results in a chiral condensate that   remains constant when the quark mass changes sign.
\end{abstract}

\maketitle

\section{Introduction}

One of the most important parameters of QCD is the total number of dynamical quarks. Asymptotic freedom is lost at the one-loop level for $N_f > 33/2$, while chiral symmetry is broken spontaneously for two or more flavors until QCD becomes conformal for a number of flavors that can be determined by means of lattice QCD simulations (see \cite{Kuti:2013} for a review). In this paper we consider one-flavor QCD, where chiral symmetry is broken explicitly by the anomaly rather than by spontaneous symmetry breaking.  As a consequence, the sign of the chiral condensate does not change when the sign of the quark mass is reversed.  On the other hand, the Banks-Casher relation \cite{Banks:1979yr} predicts that the chiral condensate does change sign when the quark mass changes sign.  The resolution of this apparent contradiction is well-known \cite{Leutwyler:1992yt,   Kanazawa:2011tt}: In the derivation of the Banks-Casher relation it is assumed that the fermionic measure (or equivalently the spectral density of the Dirac operator) is positive definite, but this assumption is violated for one-flavor QCD when the quark mass is negative.  In this paper we will show that in this case the relation between the Dirac spectrum and the chiral condensate may be determined by an alternative mechanism \cite{Osborn:2005ss}, which was first observed for QCD at nonzero baryon number chemical potential.  In the latter case the spectral density is not positive definite due to the phase of the fermion determinant, and the discontinuity of the chiral condensate when the quark mass crosses the imaginary axis does not arise from a dense spectrum of eigenvalues on the imaginary axis but rather from an oscillating spectral density in the complex plane with an amplitude that increases exponentially with the volume and a period that is inversely proportional to the volume.  An analogous mechanism is at work in other physical situations, e.g., in one-dimensional one-flavor U(1) gauge theory at nonzero chemical potential, where the Dirac spectrum is an ellipse in the complex plane while the chiral condensate only has a discontinuity across the imaginary axis \cite{Ravagli:2007rw}, in two-color QCD at nonzero chemical potential \cite{Akemann:2010tv}, or in QCD at large isospin density with mismatched quark chemical potentials \cite{Kanazawa:2014lga}.

In this paper we obtain simple analytical expressions for the one-flavor microscopic spectral density of the Dirac operator at zero $\theta$-angle.  These expressions allow us to explicitly apply the above-mentioned alternative mechanism in the case of one-flavor QCD. Analytical results for the spectral density at fixed topological charge $\nu$ are well-known \cite{Verbaarschot:1993pm,Damgaard:1997ye,Wilke:1997gf,Damgaard:1998xy} (see \cite{Verbaarschot:2000dy,Verbaarschot:2009jz} for reviews), but simple expressions for the spectral density at fixed $\theta$-angle have not yet appeared in the literature. In \cite{Damgaard:1999ij,Kanazawa:2011tt}, the spectral density at fixed $\theta$-angle was studied numerically, and several analytical results were obtained as well.  It was also realized that an analytical expression for the spectral density could be derived by combining the expression for the spectral density in terms of microscopic partition functions \cite{Akemann:1998ta} with expressions for the partition function at fixed $\theta$-angle. At the time the paper \cite{Damgaard:1999ij} was published these expressions were only known for two flavors \cite{Leutwyler:1992yt}, while an expression for the three-flavor partition function is required to obtain the one-flavor spectral density.  General results for more flavors were derived in \cite{Lenaghan:2001ur}, and they were used to obtain the spectral density at fixed $\theta$ in \cite{Kanazawa:2011tt}.  However, simpler analytical results can be obtained using identities for sums of products of Bessel functions which we derive in this paper, and as far as we know, some of these identities are not known in the literature.

The physics of one-flavor QCD was recently reviewed in \cite{Creutz:2006ts}, where some of the questions that are raised in the present paper were also addressed. However, the relation between the chiral condensate and the spectral density of the Dirac operator turns out to be much more intricate than anticipated in \cite{Creutz:2006ts}.  We show that for negative quark mass the chiral condensate results from large cancellations between the contributions of the zero and nonzero modes \cite{Kanazawa:2011tt} and that the oscillations in the spectral density are essential for the continuity of the chiral condensate.

This paper is organized as follows.  In Section~\ref{sec:relation} we discuss the relation between the spectrum of the Dirac operator and the chiral condensate for one-flavor QCD.  Also, we briefly comment on the case of several flavors.  Analytical results for the spectral density are derived in Section~\ref{sec:density}, and the condensate is evaluated in Section~\ref{sec:condensate}.  Concluding remarks are made in Section~\ref{sec:conclusions}.  In Appendix~\ref{app:uniform} we investigate under what conditions the thermodynamic limit and the sum over topological sectors can be interchanged.  In Appendix~\ref{app:sums} we derive addition theorems for products of Bessel functions.  Asymptotic results for the spectral density are worked out in Appendix~\ref{app:asympt}.  In Appendix~\ref{app:integrals} we give integrals that are used to compute the chiral condensate. Asymptotic results for the chiral condensate are worked out in Appendix~\ref{app:condensate}.  The mass independence of the chiral condensate is shown in Appendix~\ref{appd}.
 
\section{Chiral condensate and spectral density of one-flavor QCD}
\label{sec:relation}

For $m \ll 1/\Lambda_\text{QCD} \sqrt V$ the one-flavor partition
function at fixed $\theta$-angle is given by \cite{Leutwyler:1992yt}
\begin{align}
  Z(m,\theta) = e^{mV \Sigma\cos\theta}\,, 
  \label{znf1}
\end{align}
where $V$ is the volume of space-time, $m$ is a quark mass which we take to be real, and $\Sigma$ is the absolute value of the chiral condensate in the limit $m=0$ and $\theta=0$. To avoid unnecessary minus signs we define the quantity $\Sigma(m,\theta)$ (which we will also refer to as the chiral condensate) by
\begin{align}
  \Sigma(m,\theta) = -\langle \bar q q\rangle =\frac 1V \frac d{dm} \log Z(m,\theta) 
  = \Sigma \cos\theta \,.
  \label{con-sig}
\end{align}
Since $\Sigma(m,\theta)$ is independent of $m$ it does not change when $m$ crosses the imaginary axis on which the eigenvalues of the Dirac operator are located.  In terms of the Dirac eigenvalues $i\lambda_k$ (with real $\lambda_k$) $\Sigma(m)$ is given by
\begin{align}
  \Sigma(m)
  &=\frac1V\Big\langle\sum_k\frac1{i\lambda_k+m}\Big\rangle\notag\\
  &=\frac 1V \int_{-\infty}^\infty d\lambda\, 
  \frac{\rho(\lambda,m)}{i\lambda+m}\,,
  \label{eq:BC}
\end{align}
where we have not displayed the dependence on $\theta$ explicitly
because Eq.~\eqref{eq:BC} is valid not only for fixed $\theta$ but
also in sectors of fixed topological charge, i.e., as a relation
between $\Sigma_\nu(m)$ and $\rho_\nu(\lambda)$ defined below.  In the
first line of Eq.~\eqref{eq:BC}, the average is over the gauge fields
weighted by the fermion determinant. In the second line,
$\rho(\lambda,m)$ is the spectral density of the Dirac operator in the
one-flavor theory, which is symmetric with respect to $\lambda=0$ and
also includes the contributions from exact zero modes.  For a non-negative spectral density, $\Sigma(m)$ in
Eq.~\eqref{eq:BC} changes sign when the quark mass changes
sign. However, for negative quark mass the fermion determinant and
hence the spectral density is not positive definite, and we will see
below that this allows for a constant chiral condensate.

The partition function at fixed $\theta$-angle can be decomposed into 
partition functions at fixed topological charge,
\begin{align}
  Z(m,\theta) = \sum_\nu e^{i\nu\theta} Z_\nu(m)\,,
\end{align}
resulting in the decomposition 
\begin{align}
  \label{consig}
  \Sigma(m,\theta) = \frac 1{Z(m,\theta)}\sum_\nu
  e^{i\nu\theta} Z_\nu(m) \Sigma_\nu(m)  
\end{align}
of the chiral condensate with
\begin{align}
  \Sigma_\nu(m) = \frac1V\frac d{dm} \log Z_\nu(m)\,.
\end{align}
Since $Z_\nu(m)/Z(m,\theta)$ is the probability of finding a gauge-field
configuration with topological charge $\nu$, the spectral density can
be decomposed as \cite{Damgaard:1999ij}
\begin{align}
  \label{eq:rho_theta}
  \rho(\lambda,m,\theta) = \frac 1{Z(m,\theta)}\sum_\nu e^{i\nu\theta}
  Z_\nu(m) \rho_\nu(\lambda,m)\,,
\end{align}
where $\rho_\nu(\lambda,m)$ is the spectral density for gauge-field
configurations with fixed topological charge $\nu$.

The one-flavor partition function in the sector of topological charge
$\nu$ is given by
\begin{align}
  Z_\nu(m) = \frac 1{2\pi} \int_{-\pi}^\pi d\theta\, e^{-i\nu \theta }
  Z(m,\theta) = I_\nu(mV \Sigma) \,, 
\end{align}
resulting in a chiral condensate at fixed $\nu$ equal to
\begin{align}
  \Sigma_\nu(m) = \Sigma \frac {I'_\nu(mV\Sigma)}{I_\nu(mV\Sigma)}\,. 
\end{align}
In the thermodynamic limit $V\to\infty$ we thus have
\begin{align}
  \lim_{V\to\infty}\Sigma_\nu(m) = \sign(m)\Sigma\,.
\end{align}
If the thermodynamic limit and the sum over $\nu$ in Eq.~\eqref{consig} could be interchanged, the chiral condensate would be given by
\begin{align}
  \label{eq:qm1}
  \lim_{V\to\infty}\Sigma(m,\theta) \overset?= \sign(m)\Sigma\,.
\end{align}
Here and below, the question mark indicates that the result holds only under certain conditions.  In fact, \eqref{eq:qm1} contradicts Eq.~\eqref{con-sig} unless $m>0$ and $\theta=0$ or $m<0$ and $\theta=\pi$, which implies that the thermodynamic limit and the sum over $\nu$ can only be interchanged in these two cases.  Why this is so is explained in detail in Appendix~\ref{app:uniform}.

Let us now try to understand Eq.~\eqref{eq:qm1} in terms of the eigenvalue density.  The rescaled spectral density at fixed $\nu$ is defined by
\begin{align}
  \label{eq:rescaled}
  \hr_\nu(\hx,\hm) \equiv \lim_{V\to \infty}\frac 1{V\Sigma}\,\rho_\nu\left
    (\frac{\hx}{V\Sigma},\frac{\hm}{V\Sigma} \right ) , 
\end{align}
where we have introduced the dimensionless variables $\hx=\lambda V\Sigma$ and $\hm=mV\Sigma$.  If the thermodynamic limit in \eqref{eq:rescaled} is taken for \emph{fixed} $\lambda=\hx/V\Sigma$ we obtain $\hr_\nu^\nz(\hx,\hm)=1/\pi$ \cite{Verbaarschot:1993pm}, where the superscript nz (``nonzero'') indicates that we have momentarily ignored the contribution from the zero modes (which will be reinstated below). Assuming that the thermodynamic limit and the sum over $\nu$ in Eq.~\eqref{eq:rho_theta} can be interchanged, we thus obtain
\begin{align}
  \hr^\nz(\hx,\hm,\theta)\overset?=\frac1\pi\,.
  \label{rhoav}    
\end{align}
At fixed $\theta$-angle the spectral density \eqref{rhoav} then results, via Eq.~\eqref{eq:BC}, in a chiral condensate given by Eq.~\eqref{eq:qm1}.  Again, Eq.~\eqref{rhoav} and the resulting Eq.~\eqref{eq:qm1} are only correct for $m>0$ and $\theta=0$ or $m<0$ and $\theta=\pi$, i.e., if the thermodynamic limit and the sum in Eq.~\eqref{eq:rho_theta} can be interchanged (which can be shown in analogy to the arguments for Eq.~\eqref{consig} in Appendix~\ref{app:uniform}). Note that in these cases the zero modes do not contribute to the chiral condensate in the thermodynamic limit, see Eq.~\eqref{eq:zm} below (which is valid for $m>0$ and $\theta=0$).

We now set $\theta=0$ for simplicity to study only the dependence on the sign of $m$ and decompose the spectral density as
\begin{align}
  \hr(\hx,\hm,\theta=0) = \frac1\pi + \Delta\hr(\hx,\hm)\,, 
\end{align}
where the additional part also includes the contribution from the zero modes, given explicitly in Eq.~\eqref{eq:rho_zm} below. To obtain a condensate that is constant as a function of $\hm$, the contribution from the additional part of the spectral density must be equal to
\begin{align}
  \int_{-\infty}^\infty d\hx\, \frac{\Delta \hr(\hx,\hm)}{i\hx+\hm} =   2\Theta(-\hm)\,,
  \label{eq:heavi}
\end{align}
where $\Theta$ denotes the Heaviside function.  Combined with the contribution $\sign(m) \Sigma$ from the $1/\pi$ term in the spectral density this gives the correct result $\Sigma(m,\theta=0)=\Sigma$. 

The solution of Eq.~\eqref{eq:heavi} for $\Delta\hr$ is not unique.  Before discussing the case of one-flavor QCD, let us look at a simpler example.  From the Fourier decomposition of the Heaviside function, we consider the spectral density
\begin{align}
  \label{eq:heavi_ex}
  \Delta \rho(\lambda, m) = -\frac{V\Sigma}\pi \left(e^{i\lambda V\Sigma -m V\Sigma}+e^{-i\lambda V\Sigma -m V\Sigma}\right),
\end{align}
which is symmetric in $\lambda$ as in QCD.  After integration according to Eq.~\eqref{eq:heavi}, the second term gives the desired $\Theta$-function and thus the mass independence of the condensate, while the first term gives a result proportional to $\Theta(m)e^{-2mV\Sigma}$ that vanishes in the thermodynamic limit.  A similar mechanism is at work in one-flavor QCD, but there are some differences.  In the next section we derive an explicit expression for $\Delta\hr$ for this case.  In Section~\ref{sec:condensate} we show that this expression indeed satisfies Eq.~\eqref{eq:heavi}, for any value of $\hm$ and not only in the thermodynamic limit.  We will see that for $m<0$ the contributions to $\Delta\hr$ are strongly oscillating. After integration according to Eq.~\eqref{eq:heavi} we get contributions to the condensate that diverge exponentially in the thermodynamic limit, i.e., for $\hm\to-\infty$.  The desired $\Theta$-function discontinuity is obtained through cancellations of the different contributions.  We will also see that the contribution of the zero modes plays an essential role for $m<0$.

Let us briefly comment on the case of several flavors.  In that case chiral symmetry is broken spontaneously, resulting in a partition function that is dominated by Nambu-Goldstone (NG) bosons in the chiral limit. In the thermodynamic limit, the NG fields $U$ align themselves with the mass term.  After diagonalizing the NG fields the mean-field action (for degenerate quark masses and with $\theta=0$) can be written as \cite{Smilga:1998dh}
\begin{align}
  S_\text{mf}&=-\frac12mV\Sigma\tr(U+U^\dagger)\notag\\
  &=-mV\Sigma\Bigg(\sum_{i=1}^{N_f-1}\cos\phi_i+\cos\sum_{i=1}^{N_f-1}\phi_i\Bigg)\,,
\end{align}
which is to be minimized as a function of the $\phi_i$.  For even $N_f$ the minimum is obtained for $\phi_i=0$ if $m>0$ and for $\phi_i=\pi$ if $m<0$.  This results in a mass dependence of the partition function in the thermodynamic limit given by
\begin{align}
  \label{eq:Nfeven}
  Z(m) = e^{N_f |m|V \Sigma}\quad(N_f\text{ even})\,,
\end{align}
in contrast to the one-flavor partition function \eqref{znf1}.  As a consequence, the chiral condensate is discontinuous at $m=0$.  For odd $N_f$ the minimum is obtained for $\phi_i=0$ if $m>0$ and for $\phi_i=\pi(1-1/N_f)$ if $m<0$.  Hence in this case the mass dependence of the partition function is given by
\begin{align}
  Z(m) =
  \begin{cases}
    e^{N_f mV \Sigma}\,, & m>0\,,\\
    e^{N_fmV\Sigma\cos\pi(1-1/N_f)}\,, & m<0\,,
  \end{cases}
  \quad(N_f\text{ odd})
\end{align}
which agrees with \eqref{znf1} for $N_f=1$ and approaches \eqref{eq:Nfeven} for large $N_f$.  For odd $N_f>1$ the chiral condensate is also discontinuous at $m=0$.

\section{\boldmath Spectral Density at $\theta = 0$}
\label{sec:density}

Let us split up the spectral density at fixed $\nu$ into the zero-mode
contribution, the quenched nonzero-mode contribution, and the
nonzero-mode contribution due to dynamical quarks,
\begin{align}
  \label{eq:decomp}
  \rho_\nu(\lambda) = \rho^\zm_\nu(\lambda) + \rho^\q_\nu(\lambda) +
  \rho^\d_\nu(\lambda) \,, 
\end{align}
where we have suppressed the dependence on $m$ for simplicity.  In the
microscopic limit of one-flavor QCD these contributions are given by
\cite{Verbaarschot:1993pm,Damgaard:1997ye,Wilke:1997gf,Damgaard:1998xy}
\begin{align}
  \label{eq:zm_nu}
  \hr^\zm_\nu(\hx) &= {|\nu|}\delta(\hx)\,, \\
  \hr^\q_\nu(\hx)&=\frac{|\hx|}2\big[J_{\nu}^2(\hx)
  -J_{\nu+1}(\hx)J_{\nu-1}(\hx)\big]\,, \\ 
  \hr^\d_\nu(\hx,\hm) &=\frac {-|\hx|}{\hx^2+\hm^2}
  \Big [\hx J_\nu(\hx) J_{\nu+1}(\hx)\!+\!\hm \frac
    {I_{\nu+1}(\hm)}{I_\nu(\hm)}J_\nu^2(\hx)\Big].
\end{align}

\begin{figure}
  \centering
  \includegraphics{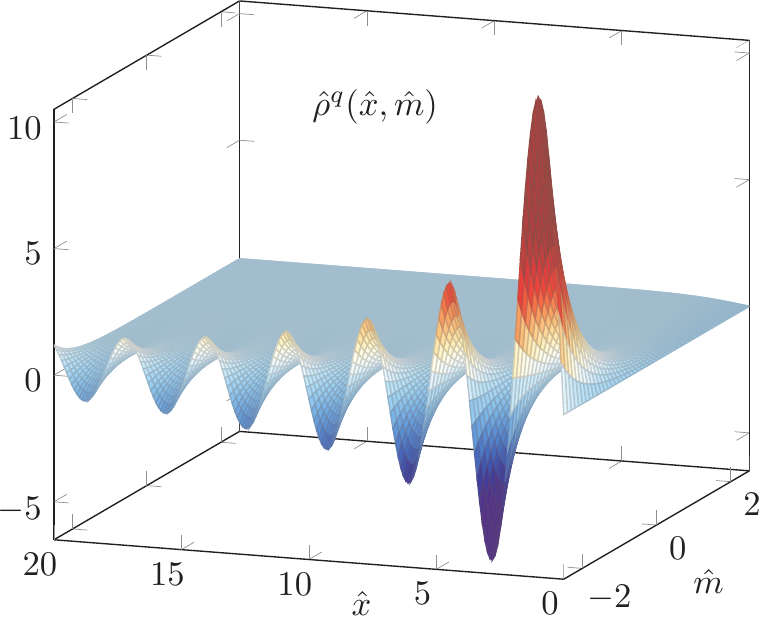}\\[5mm]
  \includegraphics{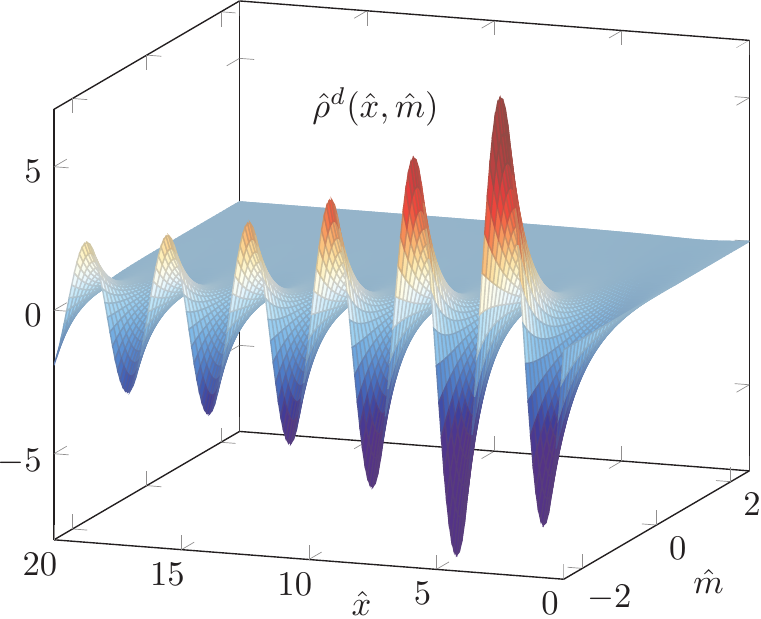}
  \caption{Three-dimensional plot of the quenched (top) and dynamical
    (bottom) part of the rescaled spectral density as a function of
    $\hx$ and $\hm$. The normalization is chosen such that the
    asymptotic value of the rescaled spectral density is equal to
    $1/\pi$.}
  \label{fig:3d}
\end{figure}

\begin{figure}
  \centerline{\includegraphics{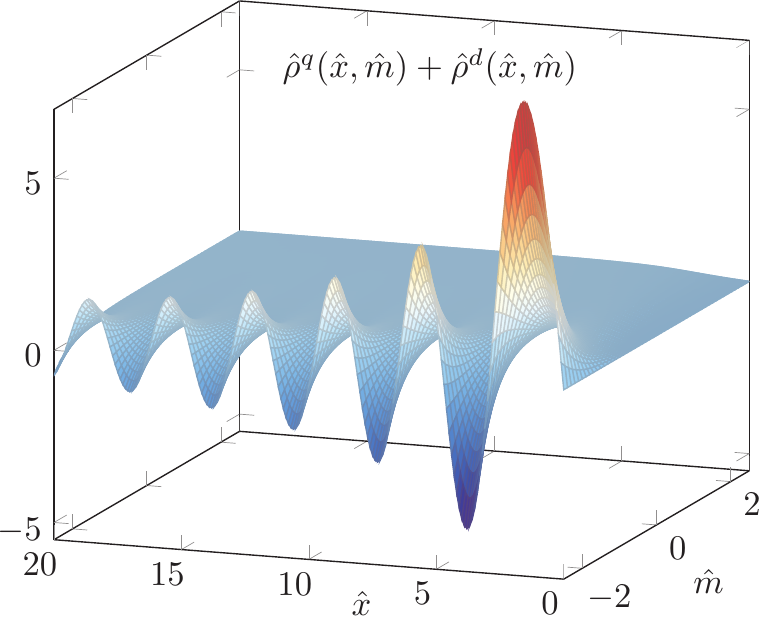}}
  \caption{Three-dimensional plot of $\hr^\q(\hx,\hm)+\hr^\d(\hx,\hm)$.}
  \label{fig:rhoqd}
\end{figure}

For the spectral density at fixed $\theta=0$ we use a decomposition
analogous to Eq.~\eqref{eq:decomp}, only with the subscript $\nu$
omitted.  For the zero-mode part of the spectral density the sum over
$\nu$ can be evaluated explicitly
\cite{Leutwyler:1992yt,Damgaard:1999ij}, resulting in
\begin{align}
  \hr^\zm(\hx,\hm) &=e^{-\hm} \sum_\nu I_\nu(\hm)|\nu| \delta(\hx)
  \notag\\
  &=e^{-\hm} \hm \big[I_0(\hm)+I_1(\hm)\big] \delta(\hx)\,. 
  \label{eq:rho_zm}
\end{align} 
The quenched part follows from the identities \eqref{eq:sum000} and
\eqref{eq:sum01-1} derived in Appendix~\ref{app:sums},
\begin{align}
  \hr^\q(\hx,\hm) &= e^{-\hm} \sum_\nu I_\nu(\hm)
  \frac{|\hx|}2\big[J_{\nu}^2(\hx) -J_{\nu+1}(\hx) J_{\nu-1}(\hx)\big]\notag \\
  &=\frac 1\pi \int_0^1 \frac {dt}{t\sqrt{1-t^2}}\, e^{-2\hm t^2} J_1(2|\hx| t)\,,
  \label{eq:rhoq} 
\end{align}
and using the identities \eqref{eq:sum001} and \eqref{eq:sum100}
the dynamical part of the spectral density is given by
\begin{align}
  \hr^\d(\hx,\hm) &= e^{-\hm} \sum_\nu I_\nu(\hm) \rho_\nu^\d(\hx,\hm)
  \notag\\ &= -\frac 2\pi \frac{|\hx|}{\hx^2+\hm^2}\int_0^1 
  \frac {dt}{\sqrt{1-t^2}}\,e^{-2\hm t^2} \notag\\
  &\qquad \times\big[ \hx t J_1(2\hx t) +\hm(1-2t^2)J_0(2\hx t)\big]\,.
  \label{eq:rhod}
\end{align}
These formulas are valid for both positive and negative quark mass.
In Fig.~\ref{fig:3d} we show three-dimensional plots of the quenched
(top) and the dynamical (bottom) part of the spectral density. The
quenched part oscillates about the asymptotic value of $1/\pi$, while
the dynamical part oscillates about zero. For negative mass the
amplitude of the oscillations increases exponentially with the volume
(i.e., with the rescaled quark mass $\hm=mV\Sigma$), while the period
in terms of $\lambda=\hx/V\Sigma$ is of order $1/V$. In
Fig.~\ref{fig:rhoqd} we plot $\hr^\q(\hx,\hm)+ \hr^\d(\hx,\hm) $.  This
figure shows that the exponentially large oscillations do not cancel,
which also follows from the asymptotic results given below.

\begin{figure}
  \centerline{\includegraphics{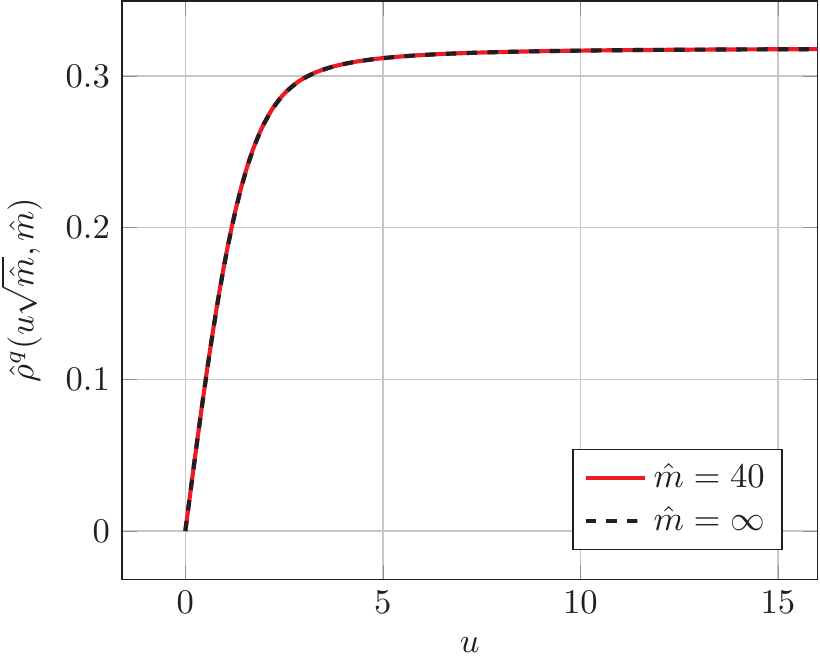}}
  \caption{Comparison of the exact result \eqref{eq:rhoq} for
    the quenched part of the spectral density for $\hm = 40$ (solid
    red curve) with the asymptotic result \eqref{as-plus} (dashed
    black curve).}
  \label{fig:dens_pos}
\end{figure}

Let us consider the large-$|\hm|$ limit of these results.  For the
zero-mode part we find
\begin{align}
  \hr^\zm(\hx,\hm)\sim
  \begin{cases}
    \displaystyle \sqrt{\frac{2\hm}\pi}\delta(\hx) \,,
    & \hm\to\infty\,,\\
    \displaystyle -\frac{e^{2|\hm|}}{\sqrt{8\pi|\hm|}}\delta(\hx) \,,
    & \hm\to-\infty\,. 
  \end{cases}
\end{align}
For the nonzero-mode parts the integrals over $t$ can be evaluated in
saddle-point approximation.  For $\hm\to\infty$ a universal scaling
function is obtained by taking the limit (see Appendix~\ref{appc})
\begin{align}
  \lim_{\hm \to \infty}\hr^\q(u \sqrt{\hm},\hm)
  &= \frac {|u| e^{-\frac{u^2}4}} {\sqrt{8 \pi}}
  \big[I_0(u^2/4)\!+\!I_1(u^2/4)\big],\label{as-plus}\\
  \lim_{\hm \to \infty}\hr^\d(u\sqrt{\hm},\hm) 
  &= -\frac {|u| e^{-\frac{u^2}4}}{\hm\sqrt{ 2\pi}} I_0(u^2/4)\,. \end{align}
This shows that the dynamical part of the spectral density is
suppressed by $1/\hm$ so that we recover the quenched result in the
large-$\hm$ limit (or, equivalently, the thermodynamic limit).  In
Fig.~\ref{fig:dens_pos} we compare the $\hm\to \infty $ limit of the
quenched part of the rescaled spectral density with the exact result
for $\hm=40$.

\begin{figure}
  \centerline{\includegraphics{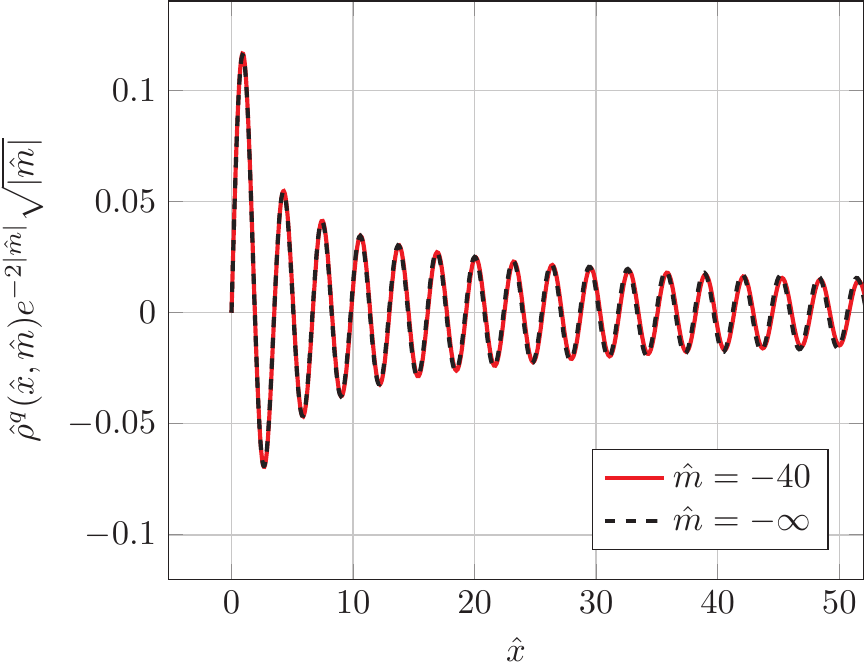}}
  \caption{Comparison of the rescaled exact result \eqref{eq:rhoq} for
    the quenched part of the spectral density for $\hm =-40$ (solid
    red curve) with the asymptotic result \eqref{as-min} (black dashed
    curve).}
  \label{fig:dens_neg}
\end{figure}

For large negative mass, the spectral density factorizes into
functions that only depend on $\hx$ or $\hm$.  A leading-order saddle
point-approximation results in (see Appendix~\ref{appa}) 
\begin{alignat}{2}
  \rho^\q(\hx,\hm) &\sim 
  \frac{e^{2|\hm|}}{\sqrt{8\pi|\hm|}}J_1(2|\hx|)\,, 
  && \hm\to-\infty\,, \label{as-min}\\
  \rho^\d(\hx,\hm) &\sim
  \frac{e^{2|\hm|}}{\sqrt{2\pi|\hm|^3}}\, |\hx| J_0(2\hx)\,,\quad
  && \hm\to-\infty\,.
\end{alignat}
In agreement with our naive expectation, also in this case the
dynamical contribution to the spectral density is suppressed by
$1/\hm$.  
The result for large negative $\hm$ increases exponentially with the
volume. In Fig.~\ref{fig:dens_neg} we show the asymptotic rescaled
result of the quenched part together with the exact result for $\hm =
-40$.

In the remainder of this section we briefly discuss an alternative
approach to compute the spectral density at fixed $\theta$-angle,
based on the expression for the one-flavor spectral density at fixed
$\nu$ derived in \cite{Akemann:1998ta}, 
\begin{align}
  \hr_\nu(\hx,\hm) = 
  \frac{(-1)^\nu}2 |\hx|(\hx^2+\hm^2)
  \frac{\hz^{N_f=3}_\nu(\hm,i\hx,i\hx)}{\hz_\nu^{N_f=1}(\hm)}\,,
\end{align}  
where we have defined $\hz(\hm)=Z(\hm/V\Sigma)$.  Using
Eq.~\eqref{eq:rho_theta} this gives \cite{Damgaard:1999ij,Kanazawa:2011tt}
\begin{align}
  \hspace*{-2mm}
  \hr(\hx,\hm,\theta) &= \frac 1{\hz(\hm,\theta)} \sum_\nu e^{i\nu\theta} 
  \hz_\nu(\hm)\hr_\nu(\hx,\hm)\notag \\
  &= \frac{|\hx|(\hx^2+\hm^2)}{2\hz(\hm,\theta)}\sum_\nu e^{i\nu\theta}(-1)^\nu
  \hz^{N_f=3}_\nu(\hm,i\hx,i\hx)\notag \\
  &= \frac{|\hx|}2(\hx^2+\hm^2)\frac {\hz^{N_f=3}(\hm,i\hx,i\hx,\theta+\pi)} 
  {\hz^{N_f=1}(\hm,\theta)} \,.
\end{align}
An integral representation of the microscopic partition function at
fixed $\theta$ was worked out in Ref.~\cite{Lenaghan:2001ur},
which gives us an explicit, though quite involved, analytical
expression for the spectral density at fixed $\theta$ \cite{Kanazawa:2011tt}.  The approach
we followed above appears to be simpler.  For the quenched
nonzero-mode part of the spectral density at $\theta=0$ it is easy to
see that the two methods lead to the same final expression. This part
can be shown to be
\begin{align}
  \hr^\q(\hx,\hm) = 
  \frac{|\hx|}2 e^{-\hm } \int_{-\pi}^\pi \frac {d\phi}{2\pi}\,
  e^{\hm \cos \phi} \hz^{N_f=2}(i\hx,i\hx,\phi+\pi)\,,
\end{align}
and using the expression for the two-flavor partition function derived
in \cite{Leutwyler:1992yt} we reproduce Eq.~\eqref{eq:rhoq}.

\section{\boldmath Chiral condensate at $\theta=0$}
\label{sec:condensate}

In this section we answer the question raised in the introduction, namely in what way a non-vanishing eigenvalue density can result in a chiral condensate that remains constant when the quark mass becomes negative.  Because of the Banks-Casher relation this is not possible for a positive definite eigenvalue density.  In essence, the discontinuity \eqref{eq:qm1} predicted by the sign-quenched theory must be canceled by another discontinuity \eqref{eq:heavi} due to the oscillating part of the spectral density.  Here we show that to obtain this discontinuity a similar mechanism is at work in one-flavor QCD as in the other cases discussed in the introduction.

We restrict ourselves to $\theta=0$, although our results can in principle be extended to nonzero $\theta$ using Eq.~\eqref{eq:S}.  The chiral condensate is related to the spectral density via Eq.~\eqref{eq:BC}.  Using the same decomposition as for the spectral density the chiral condensate can be decomposed as
\begin{align}
  \Sigma(m) = \Sigma^\zm(m) + \Sigma^\q(m) + \Sigma^\d(m)\,.
\end{align}
We define $\hs(\hm)=\Sigma(\hm/V\Sigma)/\Sigma$ to simplify the
notation in the microscopic domain.  The contribution from the zero
modes in this domain is well known \cite{Leutwyler:1992yt},
\begin{align}
  \hs^\zm(\hm) = e^{-\hm}\big[I_0(\hm) + I_1(\hm)\big]\,.
  \label{cond-zero}
\end{align}
The contributions of $\rho^\q(\hx,\hm) $ and $\rho^\d(\hx,\hm)$ to the
chiral condensate can be obtained using Eqs.~\eqref{eq:rhoq} and
\eqref{eq:rhod} and performing the integral in Eq.~\eqref{eq:BC},
which in our notation and using the symmetry of the density becomes
\begin{align}
  \hs^{\q,\d}(\hm)=2\hm \int_0^\infty d\hx\, 
  \frac{\hr^{\q,\d}(\hx,\hm)}{\hx^2+\hm^2}\,. 
\end{align}
These integrals are known analytically (see
Appendix~\ref{app:integrals}), resulting in
\begin{align}
  \label{cond-anq}
  \hs^\q(\hm) &=\frac 1{\pi\hm} \int_0^1 
  \frac {dt\, e^{-2\hm t^2}}{t^2\sqrt{1-t^2}}
  \left[ 1 - 2t|\hm| K_1(2t|\hm|)\right],\\
  \label{cond-an}
  \hs^\d(\hm) &= -\frac {4}\pi \int_0^1  \frac {dt\,t\,e^{-2\hm t^2}}
  {\sqrt{1-t^2}} \\
  &\quad\times
  \big[t\hm K_0(2t|\hm|)+(1-2t^2)|\hm|K_1(2t|\hm|)\big]\,.\notag
\end{align}

For $|\hm|\to0$ we have $\hs^\zm(\hm)\to1$, i.e., in this limit the
chiral condensate is entirely due to the zero modes.  It is
straightforward to show that both $\hs^\q(\hm)$ and $\hs^\d(\hm)$
vanish for $|\hm|\to0$.

Before giving the exact result for $\hs(\hm)$, let us look at the asymptotic behavior for $|\hm| \gg 1$.  For the zero-mode contribution we find (see Appendix~\ref{app:condensate})
\begin{align}
  \hs^\zm(\hm)\sim
  \begin{cases}
    \displaystyle \sqrt{\frac2{\pi\hm}}\,, & \hm\to\infty\,,\\[3mm]
    \displaystyle \frac{e^{2|\hm |}}{\sqrt{8\pi |\hm|^3}}\,, & \hm\to-\infty\,,
  \end{cases}
  \label{eq:zm}
\end{align}
i.e., while the contribution of the zero modes is suppressed as
$\hm\to\infty$, it grows exponentially as $ \hm\to-\infty$.  As was
already observed in \cite{Kanazawa:2011tt}, in order to get a
mass-independent chiral condensate, this exponential growth must be
canceled by the contributions of the nonzero modes.  The large-$|\hm|$
behavior of these contributions can be analyzed using a similar
approach as in Appendix~\ref{app:asympt}, and we obtain
(see Appendix~\ref{app:condensate})
\begin{align}
  \label{eq:q}
  \hs^\q(\hm) & \sim
  \begin{cases}
    \displaystyle 1-\sqrt{\frac2{\pi\hm}}+\frac 1{2\hm}\,, 
    &\hm \to \infty\,, \\[3mm]
    \displaystyle -\frac {e^{2|\hm|}}{\sqrt{8\pi |\hm|^3}}\,,
    & \hm \to -\infty\,,
  \end{cases}\\
  \hs^\d(\hm) & \sim 
  \begin{cases}
    \displaystyle -\frac 1{2\hm}\,, & \hm \to \infty\,, \\[3mm]
    \displaystyle 2 -\frac {1}{|\hm|}\,, & \hm \to -\infty\,. 
  \end{cases}
\label{asymd}
\end{align}
We see that the dynamical part is finite, while the quenched part
diverges in the thermodynamic limit for negative mass.  As already
observed in \cite{Kanazawa:2011tt}, the leading divergence in
Eq.~\eqref{eq:q} exactly cancels the leading divergence of the
zero-mode part in Eq.~\eqref{eq:zm}.  To extract a finite result for
the chiral condensate, the cancellation has to be implemented
analytically for arbitrary $\hm$. This can be achieved by observing
that the zero-mode contribution can be rewritten as
\begin{align}
  \hs^\zm(\hm) &= e^{-\hm}\big[I_0(\hm) + I_1(\hm)\big] \notag\\
  &= \frac 1{\pi\hm} \int_0^1 \frac{dt}{t^2\sqrt{1-t^2}}
  \big(1-e^{-2\hm t^2}\big)\,, 
\end{align}  
which can be checked by Mathematica \cite{mathematica}.  Adding this result to the quenched part of the chiral condensate we obtain
\begin{align}
  &\hs^\q(\hm)+\hs^\zm(\hm) = \frac 1{\pi\hm}  \int_0^1
  \frac{dt}{t^2\sqrt{1-t^2}} \notag\\
  &\qquad\qquad\qquad\quad \times
  \left [ 1 - e^{-2\hm t^2} 2t|\hm| K_1(2t|\hm|) \right].
  \label{cond-zmq}
\end{align}
The asymptotic behavior of this result is given by (see
Appendix~\ref{app:condensate})
\begin{align}
  \hs^\q(\hm)+\hs^\zm(\hm) \sim
  \begin{cases}
    \displaystyle 1+\frac 1{2\hm}\,, & \hm \to \infty\,,\\[3mm]
    \displaystyle -1+\frac 1{|\hm|}\,, & \hm \to -\infty\,.
  \end{cases}
  \label{asymq}
\end{align}
We observe that the asymptotic forms of $\hs^\q(\hm)+\hs^\zm(\hm)$ and
$\hs^\d(\hm)$ add up to one.  In fact, the relation
\begin{align}
  \hs^\zm(\hm)+\hs^\q(\hm)+\hs^\d(\hm) = 1  
\end{align}
holds for all $\hm$, which shows explicitly that the chiral condensate
is continuous when the quark mass crosses the imaginary axis.  We
prove this relation in Appendix~\ref{appd} by showing that the second
derivative with respect to $\hm$ can be expressed as an integral over
the total derivative of a function $f(t,\hm)$ that vanishes at the
endpoints of the integration domain.  

\begin{figure}[t!]
  \centerline{\includegraphics{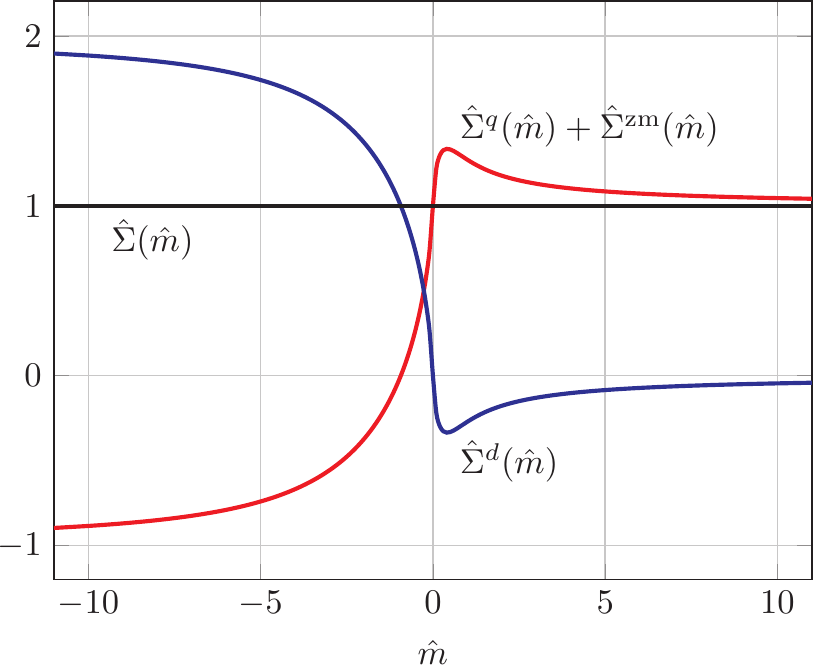}}
  \caption{The rescaled chiral condensate (black solid curve) for
    one-flavor QCD as a function of $\hm$ is the sum of a quenched
    contribution, which includes the contribution of the zero modes,
    (red curve) and a contribution due to the dynamical quarks (blue
    curve). In the thermodynamic limit both the red and the blue curve
    develop a discontinuity at $m=\hm/V\Sigma=0$.}
  \label{fig:cond}
\end{figure}

In Fig.~\ref{fig:cond} we show $\hs^\q(\hm)+\hs^\zm(\hm)$ (red),
$\hs^\d(\hm)$ (blue), and the sum of the two contributions (black),
which indeed equals one.

\section{Conclusions}
\label{sec:conclusions}

We have obtained simple analytical expressions for the microscopic spectral density of the Dirac operator for one-flavor QCD at zero $\theta$-angle.  These results enabled us to clarify the relation between the spectral density and the chiral condensate and to explain the puzzle that in the thermodynamic limit the chiral condensate develops a discontinuity in sectors of fixed topological charge, while after summing over all sectors the discontinuity disappears. The underlying reason is that for negative mass the spectral density is no longer positive definite, which invalidates the Banks-Casher relation. A different mechanism, first discovered within the context of QCD at nonzero chemical potential, takes over. The essence of this mechanism is that an oscillating part of the spectral density with a period inversely proportional to the volume and an amplitude that diverges exponentially with the volume can give rise to a contribution to the chiral condensate that is discontinuous in the thermodynamics limit at a location where there is no dense line of eigenvalues.  For QCD at nonzero chemical potential this mechanism creates a discontinuity of the chiral condensate across the imaginary axis, while the eigenvalues of the Dirac operator are scattered in a two-dimensional area around the imaginary axis.  For one-flavor QCD the discontinuity due to the oscillating part of the spectral density cancels the discontinuity of the chiral condensate of the sign-quenched theory so that the chiral condensate remains constant when the mass crosses the imaginary axis.  An important difference to QCD at nonzero chemical potential is the role played by the zero modes, which cancel a divergent contribution due to the nonzero modes. The Silver Blaze property \cite{Cohen:2003kd} of the chiral condensate could be shown by rewriting the contributions to the chiral condensate in terms of a total derivative. Remarkably, cancellations in the baryon number for QCD at nonzero chemical potential could also be explained in terms of total derivatives \cite{Greensite:2013vza}. Whether this is a coincidence or a generic feature of the Silver Blaze problem will be deferred to future work. Our results can be generalized to arbitrary $\theta$-angle and more flavors, which we also hope to address in a future publication.

\begin{acknowledgments}
  We acknowledge support by the Alexander-von-Hum\-boldt Foundation   (JV), U.S.\ DOE Grant No.\ DE-FG-88ER40388 (JV), and DFG Grant   SFB/TRR-55 (TW).  We also thank Jacques Bloch, Poul Damgaard, Takuya   Kanazawa, Mario Kieburg, and Kim Splittorff for useful discussions.
\end{acknowledgments}

\appendix

\section{Interchange of thermodynamic limit and sum over topological sectors}
\label{app:uniform}

To understand under what conditions the thermodynamic limit and the sum over $\nu$ in Eq.~\eqref{consig} can be interchanged, let us rewrite this equation in the form
\begin{align}
  \Sigma(m,\theta)&=\Sigma\cos\theta
  +\sum_\nu\frac{e^{i\nu\theta}Z_\nu(m)}{Z(m,\theta)}
  \left[\Sigma_\nu(m)-\Sigma\cos\theta\right]\notag\\
  &=\Sigma\cos\theta+\Sigma\sum_\nu f_\nu(mV\Sigma) 
  \label{eq:sum}
\end{align}
with
\begin{align}
  f_\nu(y)&= e^{i\nu\theta}e^{-y\cos\theta}
  \left[I_\nu'(y)-I_\nu(y)\cos\theta\right].
\end{align}
Using the asymptotic behavior of the Bessel functions we find that for large $|m|V\Sigma$ the terms in the sum behave like
\begin{align}
  f_\nu(mV\Sigma)\propto
  \begin{cases}
    e^{mV\Sigma(1-\cos\theta)}\,, & m>0\,,\\
    e^{|m|V\Sigma(1+\cos\theta)}\,, & m<0\,,
  \end{cases}
\end{align}
where $\propto$ means proportional up to some power of the argument. Hence the thermodynamic limit of these terms does not exist unless $m>0$ and $\theta=0$ or $m<0$ and $\theta=\pi$, which implies that the thermodynamic limit and the sum over $\nu$ cannot be interchanged unless we consider one of these two cases.  To show that they can indeed be interchanged in these cases we still need to show that the sum in Eq.~\eqref{consig} is then uniformly convergent in $mV\Sigma$.  Imposing a lower cutoff at $\nu=n$ on the sum in Eq.~\eqref{eq:sum} yields in both cases
\begin{align}
  \sign(m)\sum_{|\nu|> n} e^{-|m|V\Sigma} [I_\nu'(|m|V\Sigma)- I_\nu(|m|V\Sigma)]\,.
  \label{eq:sum1}
 \end{align}
Using recursion relations for the derivative of the Bessel function we observe that the sum in \eqref{eq:sum1} is a telescopic sum given by \begin{align}
  e^{-|m|V\Sigma} [I_{n}(|m|V\Sigma)-I_{n+1}(|m|V\Sigma)]\,.
  \label{tele}
\end{align}
One can show that \eqref{tele} is non-negative and that for large $n$ it assumes a maximum of $\sim c/n^2$ with $c=e^{-3/2}\sqrt{27/2\pi}$ at $|m|V\Sigma\sim n^2/3$.  (We have shown this by summing the asymptotic expansion \cite[Eq.~(9.7.1)]{Abramowitz:1964} of the Bessel function to all orders.)  Hence the sum in \eqref{eq:sum1} is bounded from above by $c/n^2$ independent of $mV \Sigma$, and therefore the sum \eqref{consig} is uniformly convergent in $mV\Sigma$.

\section{Addition theorems for products of Bessel functions}
\label{app:sums}

In this appendix we derive several addition theorems for products of
Bessel functions, starting from the sum
\begin{align}
  S_{a,b,c}(x,m,\theta)=\sum_{\nu=-\infty}^\infty e^{i\nu\theta}
  I_{\nu+a}(m) J_{\nu+b}(x)J_{\nu+c}(x) 
\end{align}
with $a,b,c\in\mathbb{Z}$.  We use the Fourier series technique
advocated in \cite{Martin:2008} and obtain from Eq.~(18) of that
reference
\begin{align}
  &J_{\nu+b}(x)J_{\nu+c}(x)
  =(-1)^{\nu+c} J_{\nu+b}(x)J_{-\nu-c}(x)\\
  &=(-1)^{\nu+c}\frac1\pi\int_{-\pi/2}^{\pi/2}d\psi\,J_{b-c}(2x\cos\psi)
  \cos(2\nu+b+c)\psi\,. \notag 
\end{align}
Substituting $\psi=(\phi-\pi)/2$ then leads to
\begin{align}
  &J_{\nu+b}(x)J_{\nu+c}(x)\notag\\
  &\quad=(-1)^{\frac{b-c}2}\int_{-\pi}^\pi \frac{d\phi}{2\pi}\,
  J_{c-b}\left(2x\sin\frac\phi2\right)e^{i\nu\phi}e^{i\frac{b+c}2\phi}\,.
\end{align}
In the derivation of this intermediate result we distinguished whether
$b+c$ is even or odd but obtained the same result in both cases.
Using the integral representation
\begin{align}
  I_{\nu}(m)=\int_{-\pi}^\pi \frac{d\phi}{2\pi}\,
  e^{i\nu\phi}e^{m\cos\phi} 
\end{align}
 of the modified Bessel function we find
\begin{align}
  S_{a,b,c}(x,m,\theta)&=(-1)^{\frac{b-c}2} \!\!\int_{-\pi}^\pi\!
  \frac{d\phi_1d\phi_2}{(2\pi)^2}
  e^{m\cos\phi_1}e^{ia\phi_1} e^{i\frac{b+c}2\phi_2} \notag\\ 
  &\quad\times J_{c-b}\left(2x\sin\frac{\phi_2}2\right)
  \sum_\nu e^{i\nu(\theta+\phi_1+\phi_2)}\notag\\
  &=(-1)^{\frac{b-c}2}e^{-ia\theta}\int_{-\pi}^\pi
  \frac{d\phi}{2\pi}\, e^{m\cos(\phi+\theta)} \notag \\
  &\quad\times
  J_{c-b}\left(2x\sin\frac\phi2\right) e^{i\left(\frac{b+c}2-a\right)\phi}\,.
  \label{eq:S} 
\end{align}
For $\theta=0$ some simplifications occur since $e^{m\cos\phi}$ is
even in $\phi$.  We again need to distinguish whether $b+c$ (and hence
$b-c$) is even or odd and obtain
\begin{align}
  &S_{a,b,c}(m,x,\theta=0)=\frac1{\pi}\int_0^\pi d\phi\,
  e^{m\cos\phi}J_{c-b}\left(2x\sin\frac\phi2\right)\notag\\
  &\qquad\times\begin{cases}
    (-1)^{\frac{b-c}2}\cos\left(\frac{b+c}2-a\right)\phi\,,
    & b+c \text{ even,} \\[1mm]  
    (-1)^{\frac{b-c+1}2}\sin\left(\frac{b+c}2-a\right)\phi\,,
    & b+c \text{ odd}. 
  \end{cases} 
\end{align}
Substituting $t=\sin\frac\phi2$ we can write this as an integral over
$t$ from 0 to 1.  Depending on the values of $a,b,c$ we can use
trigonometric identities to express the last factor in the integrand
in terms of $t$.  In particular, we obtain the following special cases
(all for $\theta=0$) that are needed in the main text,
\begin{align}
  &\sum_\nu I_\nu(m) J_{\nu}^2(x)=\frac 2\pi \int_0^1 \frac
  {dt}{\sqrt{1-t^2}}\, e^{m-2mt^2}J_0(2xt)\,,
  \label{eq:sum000} \\
  &\sum_\nu I_\nu(m) J_{\nu+1}(x) J_{\nu-1}(x)\notag\\
  &\qquad\qquad=-\frac 2\pi \int_0^1 
  \frac {dt}{\sqrt{1-t^2}}\,e^{m-2mt^2} J_2(2xt)\,,\label{eq:sum01-1}\\
  &\sum_\nu I_\nu(m) J_\nu(x) J_{\nu+1}(x)
  =\frac 2\pi \int_0^1 \!
  \frac {t\,dt}{\sqrt{1-t^2}} e^{m-2mt^2}\!J_1(2xt), \label{eq:sum001}\\
  &\sum_\nu  I_{\nu+1}(m) J_\nu^2(x)
  =\frac2\pi\int_0^1\frac{(1-2t^2)dt}{\sqrt{1-t^2}}\,
  e^{m-2mt^2}J_0(2xt)\,.\label{eq:sum100} 
\end{align}
These identities are related to Neumann's addition theorem for Bessel
functions, and similar identities have been discussed in the
literature.  However, we are not aware of whether all these results
are known.

The expression \eqref{eq:S} can be used to generalize our results in
the main text to arbitrary $\theta$-angle.

\section{Asymptotic behavior of the spectral density}
\label{app:asympt}

In this appendix we derive asymptotic results for the spectral density at zero $\theta$-angle.  Since all contributions to the spectral density are even in $\hx$ we restrict ourselves to $\hx\ge0$ to avoid having to write absolute values.

\subsection{\boldmath Asymptotic behavior for $\hm\to\infty$ with
  $\hx$ fixed}
\label{appb}

For $\hm\to\infty$ the integrals are dominated by the region
$t\approx0$.  Substituting $y=\sqrt{2\hm}t$ in Eq.~\eqref{eq:rhoq}
gives
\begin{align}
  \hr^\q(\hx,\hm)&=\frac 1\pi\int_0^{\sqrt{2\hm}}\frac{dy}{y\sqrt{1-{y^2}/{2\hm}}}\,
  e^{-y^2}J_1\left(\frac{2\hx y}{\sqrt{2\hm}}\right) 
\end{align}
and hence in leading order \cite{Damgaard:1999ij}
\begin{align}
  \hr^\q(\hx,\hm)\sim\frac1\pi\frac{\hx}{\sqrt{2\hm}}
  \int_0^\infty dy\,e^{-y^2}=\frac\hx{\sqrt{8\pi\hm}}\,. 
\end{align}
Substituting $y=\sqrt{2\hm}t$ in Eq.~\eqref{eq:rhod} we observe that
only the term proportional to $J_0$ contributes to the leading-order
result \cite{Damgaard:1999ij}
\begin{align}
  \hr^\d(x,m)\sim-\frac2\pi\frac\hx\hm\frac1{\sqrt{2\hm}}
  \int_0^\infty dy\,e^{-y^2}
  =-\frac\hx{\sqrt{2\pi\hm^3}}\,. 
\end{align}
We note that $\hr^\d$ is suppressed by one power of $1/\hm$ compared
to $\hr^\q$.  The leading-order expressions above are not suitable to
compute the chiral condensate since the integrals over $\hx$ diverge.

\subsection{\boldmath Asymptotic behavior for $\hm\to\infty$ and
  $\hx\sim\sqrt{\hm}$}
\label{appc}

We now substitute $y=2\hx t$ in Eq.~\eqref{eq:rhoq}, resulting in
\begin{align}
  \hr^\q(\hx,\hm)&=\frac1\pi\int_0^{2\hx}\frac{dy}{y\sqrt{1-{y^2}/{4\hx^2}}}\,
  e^{-\frac{\hm}{2\hx^2}y^2}J_1(y)\,.
\end{align}
For $\hx\sim\sqrt{\hm}$ the square root in the integrand can be approximated by one in the large-$\hm$ limit (at the upper limit, the integrand is suppressed by $e^{-2\hm}$).  Defining $\hx=2u\sqrt{\hm}$ we obtain
\begin{align}
  \hr^\q(2u\sqrt{\hm},\hm)&\sim\frac1\pi\int_0^\infty dy\,
  e^{-\frac{y^2}{8u^2}}\frac{J_1(y)}y\notag\\
  &=\frac1{\sqrt{2\pi}}ue^{-u^2}\left[I_0(u^2)+I_1(u^2)\right].
  \label{eq:82}
\end{align}
For the dynamical part of the spectral density we proceed similarly.
Substituting $y=2\hx t$ and then $\hx=2u\sqrt{\hm}$ in
Eq.~\eqref{eq:rhod} we find that also in this case only the term
proportional to $J_0$ contributes to the leading-order result
\begin{align}
  \hr^\d(2u\sqrt{\hm},\hm)&\sim-\frac1{\pi\hm}
  \int_0^\infty dy\,e^{-\frac{y^2}{8u^2}}J_0(y)\notag \\
  &=-\frac1\hm\sqrt{\frac2\pi}\,ue^{-u^2}I_0(u^2)\,.
  \label{eq:83} 
\end{align}
Hence the dynamical contributions are suppressed by $1/\hm$ also in
this limit.  In the last step of Eqs.~\eqref{eq:82} and \eqref{eq:83}
we used \cite[(6.631)]{Gradshteyn:2007}.

\subsection{\boldmath Asymptotic behavior for $\hm\to-\infty$ with
  $\hx$ fixed}
\label{appa}

For $\hm<0$ we substitute $y=2|\hm|(1-t^2)$ in Eq.~\eqref{eq:rhoq} to
obtain
\begin{align}
  \hr^\q(\hx,\hm)&=\frac{e^{2|\hm|}}{\pi\sqrt{8|\hm|}}\int_0^{2|\hm|}
  \frac{dy}{\left(1-{y}/{2|\hm|}\right)\sqrt{y}}\notag\\
  &\qquad\qquad\times e^{-y}J_1\left(2\hx\sqrt{1-\frac{y}{2|\hm|}}\right).
  \end{align}
In the large-$|\hm|$ limit at fixed $\hx$, the leading-order term is
given by
\begin{align}
  \hr^\q(\hx,\hm)&\sim\frac{e^{2|\hm|}}{\pi\sqrt{8|\hm|}}
  J_1(2\hx)\int_0^\infty\frac{dy}{\sqrt{y}}\,e^{-y} \notag \\
  &=\frac{e^{2|\hm|}}{\sqrt{8\pi|\hm|}}J_1(2\hx)\,.
\end{align}
For the next-to-leading order (NLO) term we use
\begin{align*}
  \frac1{1- y/{2|\hm|}}&\sim 1+\frac y{2|\hm|}\,,\\
  J_1\!\left(2\hx\sqrt{1-\frac{y}{2|\hm|}}\right)&\sim J_1(2\hx)
  -\frac{y}{4|\hm|}\big[2\hx J_0(2\hx)\!-\!J_1(2\hx)\big]
\end{align*}
so that
\begin{align}
  \hr^\q_\text{NLO}(\hx,\hm)&=\frac{e^{2|\hm|}}{\pi\sqrt{8|\hm|}}
  \int_0^\infty \frac{dy}{\sqrt y}\,e^{-y}
  \frac y{2|\hm|} \notag\\ 
  & \quad\times \left\{ J_1(2\hx)
    -\frac12\big[2\hx J_0(2\hx)-J_1(2\hx)\big]\right\}
    \notag\\
  &=\frac{e^{2|\hm|}}{16\sqrt{2\pi|\hm|^3}}
  \big[3J_1(2\hx)-2\hx J_0(2\hx)\big]\,. 
\end{align}
The asymptotic form of the dynamical part of the spectral density can
be derived in the same way.  Only the term proportional to $J_0$ in
Eq.~\eqref{eq:rhod} contributes to the leading-order result 
\begin{align}
  \hr^\d(\hx,\hm)&\sim \frac{e^{2|\hm|}}{\pi\sqrt{2|\hm|}}\frac   \hx{|\hm|}
J_0(2\hx)\int_0^\infty\frac{dy}{\sqrt y}\,e^{-y}\notag \\
  &=\frac{e^{2|\hm|}}{\sqrt{2\pi|\hm|^3}}\hx J_0(2\hx)\,.
\end{align}
We observe that the dynamical contribution is again suppressed by $1/\hm$ and that it is of the same order as $\hr^\q_\text{NLO}(\hx,\hm)$.

\section{Integrals over Bessel functions}
\label{app:integrals}

The following integrals over Bessel functions are known
\cite{mathematica,Gradshteyn:2007}.
\begin{align}
  \int_0^\infty dx\, \frac{J_1(2xt) }{x^2+m^2} 
  &= \frac1{2tm^2}-\frac1{|m|}K_1(2t|m|)\,,\\
  \int_0^\infty dx\, \frac{x^2J_1(2xt)}{(x^2+m^2)^2} &= tK_0(2t|m|)\,,\\
  \int_0^\infty dx\, \frac{xJ_0(2xt)}{(x^2+m^2)^2} &= \frac t{|m|}K_1(2t|m|)\,.
\end{align}
They have been used to calculate the chiral condensate from the
spectral density.

\section{Asymptotic behavior of the chiral condensate}
\label{app:condensate}

In this appendix we derive the asymptotic expansions of the three contributions to the chiral condensate given in Eqs.~\eqref{cond-zero}, \eqref{cond-anq}, and \eqref{cond-an}.

\subsection{\boldmath Asymptotic behavior for $\hm \to \infty$}

The asymptotic behavior of the zero-mode part of the chiral condensate
in Eq.~\eqref{cond-zero} simply follows from the asymptotic expansions
of the modified Bessel functions. For $\hm\to\infty$ we have
\begin{align}
  \hs^\zm(\hm) = \sqrt{\frac 2{\pi \hm}}+O(1/\hm^{3/2})\,.
\end{align}
The asymptotic expansion of the integrals \eqref{cond-anq} and
\eqref{cond-an} for $\hs^\q(\hm) $ and $\hs^\d(\hm) $ is more
complicated.  Because of the factor $e^{-2\hm t^2}$ the main
contribution to the integrals comes from the region close to
$t=0$. Neglecting subleading terms we obtain \cite{mathematica}
\begin{align}
  \hs^\q(\hm) &\sim \frac 1{\pi\hm} \int_0^\infty \frac {dt\,e^{-2\hm t^2}}{t^2}
  [1- 2t\hm K_1(2t\hm)]\notag \\
  &= \frac {\sqrt{2\hm}}{\pi\hm} \Big[-\sqrt \pi + \pi U\Big(\!-\frac 12, 0, \frac\hm2\Big)\Big]\,,
\end{align}
where $U(a,b,z)$ is a confluent hypergeometric function (a.k.a.\
Kummer's function). Using the asymptotic behavior
\cite[Eq.~(13.5.2)]{Abramowitz:1964}
\begin{align}
  U\Big(\!-\frac12, 0, \frac\hm2\Big) 
  \sim \sqrt{\frac\hm2}\Big(1+\frac1{2\hm}+\ldots\Big)
\end{align}
of this function we obtain
\begin{align}
  \hs^\q(\hm) \sim 1 - \sqrt{\frac2{\pi\hm}} + \frac 1{2\hm}\,.
\end{align}
In \eqref{cond-an} we substitute $y=2t\hm$ to obtain in leading order
\begin{align}
  \hs^\d(\hm) & \sim -\frac{4\hm}\pi \int_0^\infty
  \frac{dy\,ye^{-y^2/2\hm}}{4\hm^2}
  \left[\frac{y}{2\hm}K_0(y)+K_1(y)\right]\notag\\
  &\sim -\frac1{\pi\hm}\int_0^\infty dy\,yK_1(y)\notag\\
  &=-\frac 1{2\hm}\,.
  \label{eq:sd+}
\end{align}

\subsection{\boldmath Asymptotic behavior for $\hm \to -\infty$}

The asymptotic form of $\hs^\zm(\hm)$ again follows from the
asymptotic expansions of the Bessel functions, which for
$\hm\to-\infty$ result in
\begin{align}
  \hs^\zm(\hm) \sim \frac {e^{2|\hm|}}{\sqrt{8\pi |\hm|^3}}
  \,.
\end{align}
For $\hm \to -\infty$ the leading contribution to the integral for
$\hs^\q(\hm)$ comes from the region close to $t=1$ because of the
factor $e^{2|\hm|t^2}$.  As in Appendix~\ref{app:asympt} we substitute
$y = 2|\hm| (1-t^2)$ and obtain to leading order
\begin{align}
  \hs^\q(\hm) &\sim \frac 1{\pi \hm}
  \int_0^{\infty} \frac {dy}{4|\hm|} e^{2|\hm|-y}
  \sqrt{\frac{2|\hm|}y} \notag\\
  &=- \frac {e^{2|\hm|}}{\sqrt {8 \pi |\hm|^3}} \,. 
\end{align}
For $\hs^\d(\hm)$ we need to compute the asymptotic form to
next-to-leading order.  The presence of the factor $e^{2|\hm|t^2}$ in
\eqref{cond-an} suggests that the leading contribution to the integral
again comes from the region near $t=1$.  Performing an asymptotic
expansion of the term involving $K$-Bessel functions (with $t$ set to
$1$ in the subleading term) we obtain
\begin{align}
  \label{eq:sd1}
  \hs^\d_1(\hm)&\sim\frac{4|\hm|}\pi\int_0^1\frac{dt\,t\,e^{2|\hm|t^2}}{\sqrt{1-t^2}}
  e^{-2|\hm|t}\sqrt{\frac\pi{t|\hm|}}\notag\\
  &\qquad\qquad\times\left[\frac12(2t^2+t-1)+\frac1{16|\hm|}\right],
\end{align}
where the subscript $1$ indicates that we are currently considering
only the region near $t=1$.  In this region we substitute
$y=2|\hm|(1-t)$ and obtain to next-to-leading order
\begin{align}
  \hs^\d_1(\hm)&\sim4\sqrt{\frac{|\hm|}\pi}\int_0^{\sqrt{|\hm|}}
  \frac{dy\,\sqrt{1-y/2|\hm|}e^{-y+y^2/2|\hm|}}
  {2|\hm|\sqrt{y/|\hm|(1-y/4|\hm|)}}\notag\\
  &\qquad\qquad\quad\times
  \left[1-\frac{5y}{4|\hm|}+\frac1{16|\hm|}\right]\notag\\
  &\sim\frac2{\sqrt\pi}\int_0^\infty\frac{dy\,e^{-y}}{\sqrt y}
  \left[1+\frac{y^2}{2|\hm|}-\frac{11y}{8|\hm|}+\frac1{16|\hm|}\right]\notag\\
  &=2-\frac1{2|\hm|}\,. 
\end{align} 
We note that in Eq.~\eqref{eq:sd1} there are actually two
exponentials.  Their combination has two maxima at the boundaries of
the integration interval, i.e., at $t=0$ and $t=1$.  This suggests
that there is a contribution from the region near $t=0$ as well, which
we call $\hs^\d_0(\hm)$.  This contribution follows in analogy to the
derivation of Eq.~\eqref{eq:sd+}.  We substitute $y=2t|\hm|$ in
\eqref{cond-an} and obtain in leading order
\begin{align}
  \hs^\d_0(\hm)&\sim\frac{4|\hm|}\pi
  \!\int_0^{\sqrt{|\hm|}}\! \frac{dy\,ye^{y^2/2|\hm|}}{4|\hm|^2}
  \left[\frac{y}{2|\hm|}K_0(y)-K_1(y)\right]\notag\\
  &\sim -\frac1{\pi|\hm|}\int_0^\infty dy\,yK_1(y)\notag\\
  &=-\frac1{2|\hm|}\,.
\end{align}
Combining the two regions we thus have
\begin{align}
  \hs^\d(\hm)\sim2-\frac1{|\hm|}\,.
\end{align}

For $\hm \to -\infty$ the exponentially diverging asymptotic terms in $\hs^\zm(\hm)$ and $\hs^\q(\hm) $ cancel so that we have to take into account the subleading (non-divergent) terms. We were not able to directly derive an asymptotic form from the expression \eqref{cond-zmq}. However, in Appendix \ref{appd} we prove that the total chiral condensate is mass-independent.  In the proof we need to determine two integration constants, but to do so we only use the asymptotic behavior for large positive mass.  Therefore we are justified to use the mass independence of $\hs(\hm)$ to determine the asymptotic behavior for large negative mass, i.e.,
\begin{align}
  \hs^\zm(\hm) +\hs^\q(\hm) &= 1-\hs^\d(\hm)\notag \\  
  &\sim -1 +\frac 1{|\hm|}\,.
\end{align}

\section{Mass independence of the chiral condensate}
\label{appd}

In this appendix we show that the chiral condensate is independent of $\hm$.  From Eqs.~\eqref{cond-an} and \eqref{cond-zmq} we read off that it is given by
\begin{align}
  \hs (\hm) = \int_0^1 dt \,\big[s_1(t,\hm) + s_2(t,\hm)\big]
\end{align}
with
\begin{align}
  s_1(t,\hm) &=-\frac {4}\pi \frac {t\,e^{-2\hm t^2}}
  {\sqrt{1-t^2}}\\
  &\quad\times\big[t\hm K_0(2t|\hm|)
  +(1-2t^2)|\hm|K_1(2t|\hm|)\big]\,,\notag\\ 
  s_2(t,\hm)&=\frac 1{\pi\hm} 
  \frac{1}{t^2\sqrt{1-t^2}} 
  \big[ 1 - e^{-2\hm t^2} 2t|\hm| K_1(2t|\hm|)\big]\,.
\end{align}
It is straightforward to show that
\begin{align}
  \frac {\partial^2}{\partial\hm^2}&|\hm|[s_1(t,\hm)+s_2(t,\hm)] \notag\\
  & = \sign(m) f_1(t,\hm) - \frac{\partial}{\partial{\hm}}\hm f_2(t,\hm)
\end{align}
with total derivatives $f_1(t,\hm)$ and $f_2(t,\hm)$ given by
\begin{align}
  f_1(t,\hm) &= \frac\partial{\partial t} \left [\frac 4\pi t e^{-2\hm t^2} \sqrt{1-t^2} K_0(2t |\hm|)\right ],\\
  f_2(t,\hm) &= \frac\partial{\partial t} \left [\frac 4\pi t^2 e^{-2\hm t^2} \sqrt{1-t^2} K_1(2t |\hm|)\right ].
\end{align}
Since the terms in square brackets vanish at $t=0$ and $t=1$ we conclude that
\begin{align}
  \frac{d^2}{d\hm^2}\left[\hm\int_0^1 dt\, \big[s_1(t,\hm)+s_2(t,\hm)\big]\right]=0\,,
\end{align}
which implies
\begin{align}
  \hs(\hm)= c_0 +\frac {c_1}{\hm}\,.
\end{align}
The integration constants follow from the $\hm\to \infty $ behavior of the chiral condensate in Eqs.~\eqref{asymd} and \eqref{asymq}. This gives $c_0 =1 $ and $c_1=0$, showing that the chiral condensate does not depend on $\hm$.

\bibliography{sumnu}
\bibliographystyle{apsrev4-1}
\end{document}